\documentclass[11pt]{amsart}
\usepackage[margin=1in]{geometry}		% See geometry.pdf to learn the layout options. There are lots.
\usepackage[parfill]{parskip}			% Activate to begin paragraphs with an empty line rather than an indent
\usepackage{graphicx}
\usepackage{amssymb}
\usepackage{epstopdf}
\title{A Symmetrical Interpretation of \linebreak the Klein-Gordon Equation}
\author{Michael B. Heaney\\3182 Stelling Drive\\Palo Alto, CA 94303\\mheaney@alum.mit.edu}
\date{13 March 2013}				% Activate to display a given date or no date
\begin{document}
\maketitle
%%%%%%%%%%%%%%%%%%%%%%%%%%%%%%%%%%%%%%%%%%%%%
\begin{abstract}
This paper presents a new Symmetrical Interpretation (SI) of relativistic quantum mechanics which postulates: quantum mechanics is a theory about complete experiments, not particles; a complete experiment is maximally described by a complex transition amplitude density; and this transition amplitude density never collapses. This SI is compared to the Copenhagen Interpretation (CI) for the analysis of Einstein's bubble experiment. This SI makes several experimentally testable predictions that differ from the CI, solves one part of the measurement problem, resolves some inconsistencies of the CI, and gives intuitive explanations of some previously mysterious quantum effects.
\end{abstract}
%%%%%%%%%%%%%%%%%%%%%%%%%%%%%%%%%%%%%%%%%%%%%
\section{Introduction}
At the fifth Solvay Congress in 1927, Einstein presented a thought experiment, later known as Einstein's bubble experiment, to illustrate what he saw as a flaw in the Copenhagen Interpretation (CI) of quantum mechanics \cite{Valentini}. In his thought experiment, a single particle is released from a source, evolves freely for a time, then is captured by a detector some distance from the source. The particle is known to be localized at the source immediately before release, as shown in Figure 1(a). The CI says that, after release, the particle's probability density evolves continuously and deterministically into a progressively more delocalized distribution, up until immediately before capture, as shown in Figures 1(b) and 1(c). Immediately after capture, the particle is known to be localized at the detector, as shown in Figure 1(d). The CI says that, upon measurement at the detector, the probability density shown in Figure 1(c) undergoes an instantaneous (in all inertial reference frames), indeterministic, and time-asymmetric collapse into the different probability density shown in Figure 1(d). Einstein's bubble pops. Einstein believed this instantaneous collapse was unphysical, implying the CI was an incomplete theory. The still unresolved question of how (or if) collapse occurs is one part of the measurement problem of the CI \cite{Wheeler}. Einstein suggested that some additional mechanism, not described by the CI, was needed to make the wavefunction progressively relocalize to its final measured shape as it approached the detector. This paper presents a new Symmetrical Interpretation (SI) of relativistic quantum mechanics that has such a mechanism, solving this part of the measurement problem. This new interpretation also makes several experimentally testable predictions that differ from the CI, resolves some inconsistencies of the CI, and gives intuitive explanations of some previously mysterious quantum effects.
%%%%%%%%%%%%%%%%%%%%%%%%%%%%%%%%%%%%%%%%%%%%%%
 \begin{figure}[htbp]
\begin{center}
\includegraphics[width=5.8in]{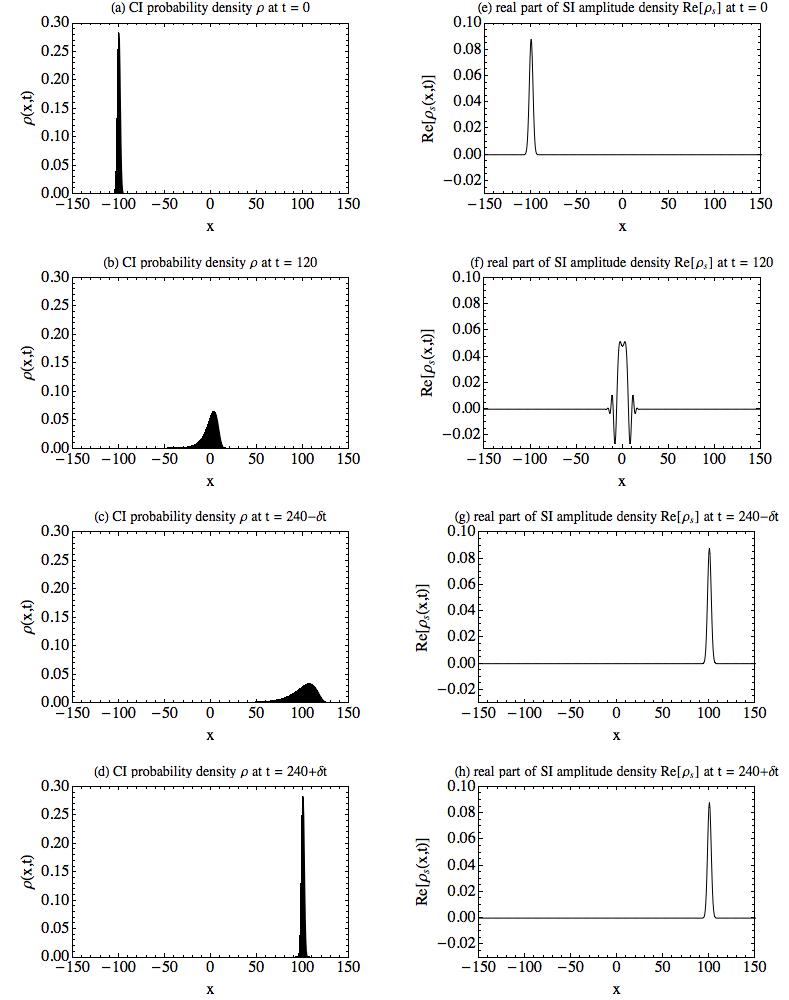}
\caption{(a-d): The Copenhagen Interpretation (CI) of Einstein's bubble. (a): The CI probability density $\rho(x,t)$ starts as a narrow distribution localized at the source. (b,c): It evolves into a broader distribution. (d): Upon measurement at $t=240$, it undergoes an instantaneous collapse into a narrow distribution localized at the detector.
(e-h): The Symmetrical Interpretation (SI) of Einstein's bubble. (e): The real part of the SI amplitude density $\rho_s(x,t)$ starts as a narrow distribution localized at the source. (f): It evolves into a broader distribution. (g): It evolves back into a narrow distribution as it approaches the detector. (h): Upon measurement at $t=240$, it does not change. The imaginary part of $\rho_s(x,t)$ behaves similarly.}
\label{default2}
\end{center}
\end{figure}
%%%%%%%%%%%%%%%%%%%%%%%%%%%%%%%%%%%%%%%%%%%%%
\section{Prior Symmetrical Interpretations} 
%%%%%%%%%%%%%%%%%%%%%%%%%%%%%%%%%%%%%%%%%%%%%
There is a long history of attempts to develop symmetrical interpretations of quantum mechanics. These attempts can be divided into two categories: symmetrical interpretations of nonrelativistic quantum mechanics, and symmetrical interpretations of relativistic quantum mechanics. Within each category are three subcategories: interpretations assuming only retarded wavefunctions, interpretations assuming a sum of retarded and advanced wavefunctions, and interpretations assuming a product of retarded and advanced wavefunctions. A retarded wavefunction has the time dependence $\psi\sim e^{-i\omega t}$ and has positive energy. The Schr\"odinger equation has only retarded solutions. An advanced wavefunction has the time dependence $\psi\sim e^{+i\omega t}$ and has negative energy. The complex conjugate of the Schr\"odinger equation has only advanced solutions.

We will first describe the prior nonrelativistic attempts. In 1964 Aharonov \textit{et al.} \cite{ABL} assumed all of the nonrelativistic CI postulates, and in addition assumed symmetrical preselected and postselected statistical ensembles. Under these assumptions, time-asymmetric wavefunction collapse still happens at every measurement, but the end results of a sequence of measurements would be time-symmetric. This showed how the time-asymmetric CI collapse postulate could give time-symmetric results under special circumstances, but it did not solve any part of the quantum measurement problem. In 1976 Davidon \cite{Davidon} proposed time-symmetric nonrelativistic postulates by assuming that the Schr\"odinger equation has both retarded and advanced solutions, and an isolated physical system was described by a product of the retarded and advanced solutions. The Schr\"odinger equation has only retarded solutions, but his proposal can be fixed by using the complex conjugate of the Schr\"odinger equation for the advanced solutions. However, Davidon postulated that the transition amplitude between any initial condition and any final condition is always equal to one, which is incorrect. Some parts of Davidon's work were later rediscovered and elaborated on by Aharonov \textit{et al.} \cite{Gruss}. However, Aharonov \textit{et al.} continued to assume the CI collapse postulate, got all of the same experimental predictions as the CI, and were still unable to solve any part of the measurement problem \cite{Aletter}. Cramer \cite{Cramer} proposed a time-symmetric quantum mechanics by assuming a particle's wavefunction was a sum of the solutions to the nonrelativistic Schr\"odinger equation and its complex conjugate. Cramer used a form of the collapse postulate, and used non-symmetric boundary conditions. Griffiths \cite{Griffiths}, Unruh \cite{Unruh}, Gell-Mann and Hartle \cite{Gellmann}, and Schulman \cite{Schulman} developed an intrinsically time-symmetric consistent histories approach. Wharton \cite{Wharton1} proposed time-symmetric nonrelativistic postulates by assuming a particle's wavefunction was a sum of the solutions to the Schr\"odinger equation and its complex conjugate, with symmetric boundary conditions. Sutherland \cite{Sutherland} assumed a product of the retarded and advanced solutions of the Schr\"odinger equation and its complex conjugate, interpreting the real part of this product as an objective density (not a probability distribution) of the mass, charge, \textit{etc.} of the particle. Sutherland \cite{Sutherland} also developed a symmetrical generalization of Bohm's nonrelativistic hidden variable theory, again using the real part of the product of the retarded and advanced solutions.

Relatively little prior work has been done on symmetrical interpretations of relativistic quantum mechanics. Griffiths \cite{Griffiths2}, Omnes \cite{Omnes}, and Blencowe \cite{Blencowe} extended the consistent histories approach to the relativistic domain. Hoyle and Narlikar \cite{Hoyle} and Davies \cite{Davies} proposed a time-symmetric relativistic quantum electrodynamics by assuming a particle was described by a sum of retarded and advanced wavefunctions. Wharton \cite{Wharton2} proposed symmetrical relativistic postulates by assuming a particle was described by a sum of the retarded and advanced wavefunction solutions to the Klein-Gordon equation, and eliminating the collapse postulate. Sutherland \cite{Sutherland} extended his symmetrical model to the relativistic domain, continuing to interpret the real part of the product as an objective density (not a probability distribution) of the mass, charge, \textit{etc.} of a particle. Sutherland \cite{Sutherland} also developed a symmetrical relativistic generalization of the Bohm model, again using only the real part of the product of the retarded and advanced solutions.

This paper proposes a new Symmetrical Interpretation (SI) of relativistic quantum mechanics, different from the CI and all of the prior symmetrical interpretations described above. This new SI postulates that quantum mechanics is a theory which describes a \textquotedblleft complete experiment," defined as an isolated, individual physical system that starts with maximally specified initial conditions, evolves in space-time, then ends with maximally specified final conditions. This is qualitatively different from the CI and these prior symmetrical interpretations, which all implicitly assume that quantum mechanics is a theory which describes an isolated, individual physical system such as a particle or group of particles. This new SI also postulates that a \textquotedblleft complete experiment" is maximally described by a complex transition amplitude density. This transition amplitude density is composed of the retarded and advanced wavefunction solutions of a relativistic wave equation, where the retarded solution satisfies only the initial conditions, and the advanced solution satisfies only the final conditions. This new SI also postulates that this transition amplitude density never undergoes collapse. This new SI will be explained, and compared to the CI, in an analysis of Einstein's bubble experiment. 
%%%%%%%%%%%%%%%%%%%%%%%%%%%%%%%%%%%%%%%%%%%%%
\section{The Copenhagen Interpretation of Einstein's Bubble Experiment}
%%%%%%%%%%%%%%%%%%%%%%%%%%%%%%%%%%%%%%%%%%%%%
Let us do a quantitative analysis of Einstein's bubble experiment, using the CI postulates as presented in standard textbooks \cite{CTDL}. The CI implicitly assumes that quantum mechanics is a theory which describes an isolated, individual physical system. The CI wavefunction postulate says an isolated, individual physical system is maximally described by a wavefunction $\Theta(\vec{r},t)$, with specified initial conditions. For experiments such as Einstein's bubble experiment, the CI assumes the isolated, individual physical system is the particle. The CI evolution postulate says the wavefunction of a free, spin-0 particle of mass $m$ will evolve continuously and deterministically according to the relativistic Klein-Gordon equation (KGE):

\begin{equation}
\left(\frac{1}{c^2}\frac{\partial^2}{\partial t^2}-\nabla^2+\frac{m^2c^2}{\hbar^2}\right)\Theta=0,
\label{ }
\end{equation}

where $c$ is the speed of light and $2\pi\hbar$ is Planck's constant.

The KGE has both retarded and advanced solutions, and both are needed for a complete set of solutions. The CI therefore assumes the general solution $\Theta(\vec{r},t)$ of the KGE is a sum of a positive energy, retarded wavefunction $\Psi(\vec{r},t)$ and a negative energy, advanced wavefunction $\Phi^\ast(\vec{r},t)$:

\begin{equation}
\Theta = \Psi + \Phi^\ast,
\label{ }
\end{equation}

where $\Theta(\vec{r},t)$ is a shorthand notation for $\Theta(x,y,z,t)$.

The complex conjugate of the KGE is:

\begin{equation}
\left(\frac{1}{c^2}\frac{\partial^2}{\partial t^2}-\nabla^2+\frac{m^2c^2}{\hbar^2}\right)\Theta^\ast=0.
\label{ }
\end{equation}

If we multiply Eq.(1) on the left by $\Theta^\ast$, multiply Eq.(3) on the left by $\Theta$, take the difference of the two resulting equations, and rearrange terms, we get:

\begin{equation}
\frac{\partial}{\partial t}\left[\frac{i\hbar}{2mc^2}\left(\Theta^\ast\frac{\partial\Theta}{\partial t}-\Theta\frac{\partial\Theta^\ast}{\partial t}\right)\right]+\nabla\cdot\left[ \frac{\hbar}{2mi}\left(\Theta^\ast\nabla\Theta-\Theta\nabla\Theta^\ast \right)\right]=0.
\label{ }
\end{equation}

The CI defines $\rho(\vec{r},t)$ as:

\begin{equation}
\rho\equiv \frac{i\hbar}{2mc^2}\left(\Theta^\ast \frac{\partial\Theta}{\partial t}-\Theta \frac{\partial\Theta^\ast}{\partial t} \right),
\label{ }
\end{equation}

and $\vec{j}(\vec{r},t)$ as:

\begin{equation}
\vec{j}\equiv\frac{\hbar}{2mi}\left(\Theta^\ast \nabla\Theta - \Theta \nabla\Theta^\ast \right),
\label{ }
\end{equation}

to get a CI local conservation law:

\begin{equation}
\frac{\partial\rho}{\partial t}+\nabla\cdot\vec{j}=0.
\label{ }
\end{equation}

The CI originally interpreted the quantity $\rho(\vec{r},t)$ as the probability density to find the particle, and $\vec{j}(\vec{r},t)$ as the probability density current. This leads to serious problems, which will be discussed later. To avoid these problems for now, we will assume $\Theta(\vec{r},t)=\Psi(\vec{r},t)$. The CI interprets $\Psi(\vec{r},t)$ as the positive energy, retarded wavefunction of a particle. (Note that we could have instead assumed $\Theta(\vec{r},t)=\Phi^\ast(\vec{r},t)$. The CI reinterprets $\Phi^\ast(\vec{r},t)$ as the positive energy, retarded wavefunction of an antiparticle.) If we assume $\Psi(\vec{r},t)$ is normalized and goes to zero at $\vec{r}=\pm\infty$, and integrate the CI local conservation law over all space $V$, we get a CI global conservation law:

\begin{equation}
\iiint_{-\infty}^{+\infty}dV\rho(\vec{r},t) = 1,
\label{ }
\end{equation}

where $dV\equiv dxdydz$ is the differential volume element. This predicts that the volume under the curve $\rho(\vec{r},t)$ is conserved and equal to one. This is the probability of finding the particle somewhere in space.

Now consider the CI of Einstein's bubble experiment shown in Figure 1(a,b,c,d). Let us use natural units and assume the following: a one-dimensional experiment; a particle mass $m\equiv1$; a positive energy, retarded wavefunction solution $\Psi(\vec{r},t)$ to the KGE; an initial normalized CI probability density $\rho(\vec{r},t_i)$ at the source that is a gaussian with mean $x_i\equiv-100$, standard deviation $\sigma_i\equiv2^{3/2}$, and momentum $p_i\equiv1.5$; and a final normalized CI probability density $\rho(\vec{r},t_f)$ at the detector that is a gaussian with mean $x_i\equiv100$, standard deviation $\sigma_i\equiv2^{3/2}$, and momentum $p_f\equiv1.5$. Figure 1(a) shows the particle's normalized CI probability density at the initial time $t_i\equiv0$ \cite{Thaller}. It is localized at the source and symmetrical in space. Figure 1(b) shows the CI probability density at $t_m\equiv120$, the midway time between release and capture. The CI probability density becomes progressively more delocalized and asymmetrical between $t_i$ and $t_m$. The asymmetry is due to the limiting speed of light. Figure 1(c) shows the CI probability density at $t_f-\delta t$, just before measurement at the detector at $t_f=240$. The CI probability density has become progressively more delocalized and asymmetrical between $t_m$ and $t_f-\delta t$. Between $t_i$ and $t_f-\delta t$, the CI probability density evolves continuously and deterministically according to the KGE. Numerical calculations confirm that the area under the curve $\rho(x,t)$ is conserved and equal to one.

The CI measurement postulate says that, upon measurement at time $t_f\equiv 240$, the amplitude $A$ for a particle having the initial wavefunction $\Theta(\vec{r},t_f-\delta t)$ to be found having the final wavefunction $\Theta(\vec{r},t_f+\delta t)$ is:

\begin{equation}
A\equiv \frac{i\hbar}{2mc^2} \iiint_{-\infty}^{+\infty}dV\left[\Theta^\ast(\vec{r},t_f+\delta t)\frac{\partial\Theta(\vec{r},t_f-\delta t)}{\partial t}-\Theta(\vec{r},t_f-\delta t)\frac{\partial\Theta^\ast(\vec{r},t_f+\delta t)}{\partial t} \right].
\label{ }
\end{equation}

$A$ is known as the CI transition amplitude. For the specific experiment shown in Figure 1(a,b,c,d), numerical calculations predict $A=0.43+0.08i$.

The CI says the probability $P$ for this transition is:

\begin{equation}
P\equiv A^\ast A.
\label{ }
\end{equation}

$P$ is known as the CI transition probability. $P$ is the conditional probability that a particle with the given initial conditions will later be found with the given final conditions. For the specific experiment shown in Figure 1(a,b,c,d), numerical calculations predict $P=0.19$.

The CI collapse postulate says that, upon measurement at $t_f$, the wavefunction $\Psi(\vec{r},t)$, and therefore the CI probability density of Figure 1(c), undergoes an instantaneous (in all inertial reference frames), indeterministic, and time-asymmetric collapse into the CI probability density of Figure 1(d). The final CI probability density is localized at the detector and symmetric in space. The CI reinterpretation of the KGE as a quantum field equation gives no further explanation of how this wavefunction collapse occurs \cite{Nikolic}. 
%%%%%%%%%%%%%%%%%%%%%%%%%%%%%%%%%%%%%%%%%%%%%
\section{The Symmetrical Interpretation of Einstein's Bubble Experiment} 
%%%%%%%%%%%%%%%%%%%%%%%%%%%%%%%%%%%%%%%%%%%%%
Let us reconsider the assumptions and postulates of the CI, and look for more symmetrical alternatives. The CI implicitly assumes that quantum mechanics is a theory which describes an isolated, individual physical system. This isolated, individual physical system is assumed to be a particle or a group of particles. In contrast, Feynman proposed that quantum mechanics is a theory which describes a \textquotedblleft complete experiment," defined as an isolated, individual physical system that starts with maximally specified initial conditions, evolves in space-time, then ends with maximally specified final conditions \cite{Feynman}. Feynman's proposal implies that a \textquotedblleft complete experiment" always has maximally specified final conditions (even though we may not know those final conditions at the start of the experiment), and that the initial and final conditions can be specified independently. Let us take Feynman's proposal as the first postulate of our Symmetrical Interpretation (SI).

Let us propose a SI amplitude density postulate which says a \textquotedblleft complete experiment" (as defined above) is maximally described by a complex transition amplitude, composed of two wavefunctions: a positive energy, retarded wavefunction $\Psi(\vec{r},t)$, which satisfies only the maximally specified initial conditions; and a negative energy, advanced wavefunction $\Phi^\ast(\vec{r},t)$, whose complex conjugate $\Phi(\vec{r},t)$ satisfies only the maximally specified final conditions. For simplicity, we will assume that the advanced wavefunction at $t_f$ is identical to the complex conjugate of the retarded wavefunction at $t_i$, but with the center translated from $(\vec{r}_i,t_i)$ to $(\vec{r}_f,t_f)$.

Let us propose SI evolution postulates for a free, spin-0 particle of mass $m$ which say the positive energy, retarded wavefunction $\Psi(\vec{r},t)$ evolves from maximally specified initial conditions according to the KGE:

\begin{equation}
\left(\frac{1}{c^2}\frac{\partial^2}{\partial t^2}-\nabla^2+\frac{m^2c^2}{\hbar^2}\right)\Psi=0,
\label{ }
\end{equation}

while the negative energy, advanced wavefunction $\Phi^\ast(\vec{r},t)$ evolves from maximally specified final conditions according to the complex conjugate of the KGE:

\begin{equation}
\left(\frac{1}{c^2}\frac{\partial^2}{\partial t^2}-\nabla^2+\frac{m^2c^2}{\hbar^2}\right)\Phi^\ast=0.
\label{ }
\end{equation}

If we multiply Eq.(11) on the left by $\Phi^\ast$, multiply Eq.(12) on the left by $\Psi$, take the difference of the two resulting equations, and rearrange terms, we get:

\begin{equation}
\frac{\partial}{\partial t}\left[\frac{i\hbar}{2mc^2}\left(\Phi^\ast\frac{\partial\Psi}{\partial t}-\Psi\frac{\partial\Phi^\ast}{\partial t}\right)\right]+\nabla\cdot\left[ \frac{\hbar}{2mi}\left(\Phi^\ast\nabla\Psi-\Psi\nabla\Phi^\ast \right)\right]=0.
\label{ }
\end{equation}

Now we will define $\rho_s(\vec{r},t)$ as:

\begin{equation}
\rho_s\equiv\frac{i\hbar}{2mc^2}\left(\Phi^\ast\frac{\partial\Psi}{\partial t}-\Psi\frac{\partial\Phi^\ast}{\partial t}\right),
\label{ }
\end{equation}

and define $\vec{j_s}(\vec{r},t)$ as:

\begin{equation}
\vec{j}_s\equiv\frac{\hbar}{2mi}\left(\Phi^\ast\nabla\Psi-\Psi\nabla\Phi^\ast \right),
\label{ }
\end{equation}

to get a SI local conservation law:

\begin{equation}
\frac{\partial\rho_s}{\partial t}+\nabla\cdot\vec{j}_s=0.
\label{ }
\end{equation}

Note that both $\rho_s(\vec{r},t)$ and $\vec{j_s}(\vec{r},t)$ are generally complex functions, and therefore cannot be interpreted as a real probability density and probability density current. Instead, we will interpret $\rho_s(\vec{r},t)$ as the amplitude density for a \textquotedblleft complete experiment," defined as an isolated, individual physical system that starts with maximally specified initial conditions, evolves in space-time, then ends with maximally specified final conditions. The complex function $\rho_s(\vec{r},t)$ will hereafter be referred to as the SI amplitude density. We will also interpret $\vec{j_s}(\vec{r},t)$ as the complete experiment amplitude current density. The complex function $j_s(\vec{r},t)$ will hereafter be referred to as the SI amplitude current density. We will also interpret the complex conjugates of these two functions as describing a complete experiment with an antiparticle.

If we assume the wavefunctions $\Psi(\vec{r},t)$ and $\Phi^\ast(\vec{r},t)$ are normalized and go to zero at $\vec{r}=\pm\infty$, and integrate the SI local conservation law over all space $V$, we get a SI global conservation law:

\begin{equation}
A_s \equiv \iiint_{-\infty}^{+\infty}dV\rho_s(\vec{r},t),
\label{ }
\end{equation}

where the SI interprets $A_s$ as the amplitude for an isolated, individual physical system to start with maximally specified initial conditions, evolve in space-time, then end with maximally specified final conditions (a \textquotedblleft complete experiment"). The complex number $A_s$ will hereafter be referred to as the SI transition amplitude. This predicts that the volume under the curve of the real part of $\rho_s(x,t)$ is conserved, and the volume under the curve of the imaginary part of $\rho_s(x,t)$ is also conserved. 

Now consider the SI of Einstein's bubble experiment shown in Figure 1(e,f,g,h). For this case, the complete experiment is: a particle is initially localized at position $x_i$ at time $t_i$, it evolves freely in space-time, then it is finally found to be localized at position $x_f$ at time $t_f$. Let us use natural units and assume the following: a one-dimensional experiment; a particle mass $m\equiv1$; an initial normalized probability density for the retarded wavefunction $\Psi(\vec{r},t_i)$ at the source that is a gaussian with mean $x_i\equiv-100$, standard deviation $\sigma_i\equiv2^{3/2}$, and momentum $p_i\equiv1.5$; and a final normalized (to -1) probability density for the advanced solution $\Phi^\ast(\vec{r},t_f)$ at the detector that is a gaussian with mean $x_f\equiv100$, standard deviation $\sigma_f\equiv2^{3/2}$, and momentum $p_f\equiv-1.5$. Figure 1(e) shows the real part of the SI amplitude density, $Re[\rho_s(x,t)]$, at the initial time $t_i\equiv0$. It is localized at the source. Figure 1(f) shows $Re[\rho_s(x,t)]$ at $t_m=120$, the midway time between release and capture. The distribution of $Re[\rho_s(x,t)]$ becomes progressively more delocalized between $t_i$ and $t_m$. Figure 1(g) shows $Re[\rho_s(x,t)]$ at $t_f-\delta t$, just before measurement at the detector. The distribution of $Re[\rho_s(x,t)]$ progressively relocalizes to the final condition between $t_m$ and $t_f-\delta t$. Numerical calculations confirm that the area under the curve of $Re[\rho_s(x,t)]$ is conserved, and the area under the curve of the imaginary part of the SI amplitude density, $Im[\rho_s(x,t)]$, is also conserved. For the specific experiment shown in Figure 1(e,f,g,h), numerical calculations predict an SI transition amplitude $A_s=0.43+0.08i$. This agrees with the calculated CI transition amplitude.

The SI complete experiment probability $P_s$ can then be defined by:

\begin{equation}
P_s\equiv A_s^\ast A_s.
\label{ }
\end{equation}

The SI interprets $P_s$ as the probability that an isolated, individual physical system will start with a given set of maximally specified initial conditions, evolve in space-time, then end with a given set of maximally specified final conditions. This is the conditional probability that a particle with the given initial conditions will later be found with the given final conditions. Since $P_s$ is time-independent, it is also the probability that a particle with the given final conditions would have been found earlier with the given initial conditions. This time symmetry is generally true for quantum transition probabilities. The positive real number $P_s$ will hereafter be referred to as the SI transition probability. For the specific experiment shown in Figure 1(e,f,g,h), numerical calculations predict an SI transition probability $P_s=0.19$. This agrees with the numerically calculated CI transition probability. 

Note that:

\begin{equation}
P_s \equiv \iiint_{-\infty}^{+\infty}dV\rho^\ast_s(\vec{r},t) \iiint_{-\infty}^{+\infty}dV\rho_s(\vec{r},t) \neq \iiint_{-\infty}^{+\infty}dV\rho^\ast_s(\vec{r},t)\rho_s(\vec{r},t),
\label{ }
\end{equation}

which implies the SI does not predict the probability density for a particle to be found at any time between the initial conditions and the final conditions of a complete experiment. This is why $\rho^\ast_s\rho_s$ is not plotted in Figure 1(e,f,g,h). This is consistent with the SI postulate that quantum mechanics is a theory which describes complete experiments, not isolated particles. 

Let us postulate that the SI amplitude density $\rho_s(\vec{r},t)$ never undergoes collapse. This is reasonable because the SI amplitude density is a quantum transition amplitude, and (as in the CI) quantum transition amplitudes do not collapse. Figure 1(f) shows $Re[\rho_s(x,t-\delta t)]$ immediately before measurement, and Figure 1(g) shows $Re[\rho_s(x,t+\delta t)]$ immediately after measurement. They are identical, in the limit $\delta t\rightarrow 0$. The same is true for $Im[\rho_s(x,t-\delta t)]$ and $Im[\rho_s(x,t+\delta t)]$. The SI amplitude density $\rho_s(\vec{r},t)$ evolves continuously and deterministically at all times. 

Note that the SI proposed in this paper is not a relativistic version of Aharonov and Vaidman's Two-State Vector Formalism (TSVF) \cite{Gruss}. The postulates are fundamentally different, and some of the experimental predictions are different. First, the TSVF assumes quantum mechanics is a theory about an isolated, individual physical system, defined as a particle or a group of particles; the SI assumes quantum mechanics is a theory about a \textquotedblleft complete experiment," defined as an isolated, individual physical system that starts with defined initial conditions, evolves in space-time, then ends with defined final conditions. Second, the TSVF assumes a nonrelativistic physical system at time \textit{t} is maximally described by a two-state vector $\langle\Phi\vert\thickspace\vert\Psi\rangle$, which is not a scalar product; the SI assumes a nonrelativistic complete experiment at time \textit{t} is maximally described by a complex amplitude density $\Phi^\ast\Psi$, which is an algebraic product. Third, the TSVF assumes wavefunctions instantaneously collapse upon measurement; the SI assumes the complex amplitude density $\Phi^\ast\Psi$ never collapses. Fourth, the TSVF gives experimental predictions that always agree with CI predictions \cite{Gruss}; the SI gives experimental predictions that sometimes disagree with CI predictions.

All of the experimentally verifiable predictions of the SI, as presented above, are formulated in terms of transition probabilities. All of the experimentally verifiable predictions of the CI can also be formulated in terms of transition probabilities \cite{Dirac}. From this perspective, the SI should have the same predictive power as the CI. Also, this suggests the use of wavefunctions to define initial and final conditions is as consistent with the SI as it is with the CI. However, this does not imply that all of the experimentally verifiable predictions of the SI and the CI are the same, as will be discussed next.
%%%%%%%%%%%%%%%%%%%%%%%%%%%%%%%%%%%%%%%%%%%%%
\section{Experimental Tests to Distinguish the Two Interpretations} 
%%%%%%%%%%%%%%%%%%%%%%%%%%%%%%%%%%%%%%%%%%%%%
For the CI general solution to the KGE given by Eq.(2), the CI probability density $\rho(\vec{r},t)$ given by Eq.(5) will contain terms that oscillate at a frequency $\omega\approx 2mc^2/\hbar$, due to interference between the positive and negative energy terms. Even if one assumes the particle starts with only a positive energy term, interactions with an electromagnetic field will add negative energy terms. The same behavior occurs for the general solution to the Dirac equation for an electron, where $\omega\approx 10^{21}Hz$. Schr\"odinger discovered the possibility of this rapid oscillating motion in 1930, naming it \textit{Zitterbewegung} \cite{Schroedinger}. This prediction of the CI is inconsistent with Newton's first law, since it implies a free particle does not move with a constant velocity. In the SI, the behavior of a free particle between the initial and final conditions is always described by the SI amplitude density $\rho_s(\vec{r},t)$ given by Eq.(14), where $\Psi(\vec{r},t)$ is always a purely positive energy wavefunction and $\Phi^\ast(\vec{r},t)$ is always a purely negative energy wavefunction. The \textit{Zitterbewegung} terms cancel each other in the SI amplitude density $\rho_s(\vec{r},t)$ and the SI amplitude current density $\vec{j_s}(\vec{r},t)$. This implies the position of a free spin-0 particle will never oscillate at a frequency $\omega\approx 2mc^2/\hbar$, so \textit{Zitterbewegung} will never occur. This prediction of the SI is consistent with Newton's first law: a free particle will always move with a constant velocity. Direct measurements of \textit{Zitterbewegung} are beyond the capability of current technology, but future technological developments should allow measurements to confirm or deny its existence, thereby distinguishing between the two interpretations.

For the CI general solution to the KGE given by Eq.(2), the energy, probability density, and probability density current can each be negative. Even if one assumes only positive energy solutions to the KGE, the probability density current can still be negative in some space-time regions \cite{Nikolic}. The CI predictions of negative probability densities and negative probability density currents are unphysical: how can a particle have a probability of less than zero of being found somewhere or of going somewhere? The CI prediction of negative energy wavefunctions makes matter unstable: a particle in a positive energy level could spontaneously decay to a negative energy level, then continue cascading to lower energy levels forever. These results led the CI to the conclusion that the KGE is not a valid single-particle quantum wave equation. These results also make the nonrelativistic limit of the KGE inconsistent with the Schr\"odinger equation. For the SI general solutions given by Eqs.(14) and (15), the facts that $\rho_s(\vec{r},t)$ and $\vec{j}_s(\vec{r},t)$ can be negative or complex are not unphysical, since the SI interprets these quantities as amplitude densities, not probability densities. In the SI, the existence of negative energy wavefunctions does not make matter unstable, since every particle is, in some sense, a product of a negative energy wavefunction and a positive energy wavefunction. This implies it is not possible to decay from a positive energy level to a negative energy level. This allows the SI to interpret the KGE as a valid single-particle quantum wave equation, in the limit where particle creation and destruction are negligible. This also makes the nonrelativistic limit of the KGE consistent with a symmetrical interpretation of the Schr\"odinger equation. Experimental tests of the validity of the KGE for complete experiments with single particles in the nonrelativistic limit, and its consistency with the Schr\"odinger equation predictions, would distinguish between the two interpretations.

The CI predicts that the wavefunction of a single particle can instantaneously, in all reference frames, collapse into a very different wavefunction. The SI predicts that the SI transition amplitude density $\rho_s(\vec{r},t)$ never undergoes collapse. An experimental test of whether collapse occurs or not would distinguish between the two interpretations.

For the CI of Einstein's bubble experiment, Figure 1(b) shows an asymmetrical CI probability density at the midpoint between the initial and final conditions. For the SI of Einstein's bubble experiment, Figure 1(f) shows a symmetrical real part of the SI amplitude density at the midpoint between the initial and final conditions. The imaginary part, not shown, is also symmetrical at the midpoint between the initial and final conditions. This difference may have experimentally testable consequences that would distinguish between the two interpretations.
%%%%%%%%%%%%%%%%%%%%%%%%%%%%%%%%%%%%%%%%%%%%%
\section{General Implications of the Symmetrical Interpretation} 
%%%%%%%%%%%%%%%%%%%%%%%%%%%%%%%%%%%%%%%%%%%%%
The SI amplitude density $\rho_s(\vec{r},t)$ never undergoes collapse, which implies that measurements and observers play no special role in the SI, as they do in the CI \cite{vonNeumann}.

The CI assumes an isolated, individual physical system is maximally described by a retarded wavefunction and maximally specified initial conditions. The SI assumes a complete experiment is maximally described by the SI amplitude density $\rho_s(\vec{r},t)$, which is composed of a retarded wavefunction and an advanced wavefunction, and maximally specified initial and final conditions. The existence of both retarded and advanced wavefunctions in the SI does not imply that particles can travel at superluminal speeds. In the SI every particle is represented by algebraic products of an advanced wavefunction and a retarded wavefunction, so the particle cannot travel to space-time locations that the retarded wavefunction cannot reach, and the KGE limits the velocity of the retarded wavefunction to less than $c$. Conversely, the particle cannot travel to space-time locations that the advanced wavefunction cannot reach. This is a type of symmetrical forward and backward causality: what happens during an experiment depends on what happened at the start of the experiment, and what will happen at the end of the experiment. This suggests the past, present, and future have equal status. This is implicit in the SI postulates, and consistent with the block universe view \cite{Price} and with the special theory of relativity. 

This property of the SI allows new and intuitive explanations for Renninger negative-result experiments (also known as \textquotedblleft null" or \textquotedblleft interaction-free" measurements) \cite{IFM}, and \textquotedblleft delayed-choice" experiments \cite{DC}. In the simplest case of these experiments, a particle has two possible paths to travel from the source to the detector. The particle seems to know if one of those paths is (or will be) blocked or open, well before it has had time to reach the block. From the SI perspective, the particle can only go to places where both the retarded and advanced wavefunctions are nonzero. A block in any space-time volume between the source and detector can prevent the advanced wavefunction from reaching the source at the time of emission, effectively letting the particle know that path is (or will be) blocked. Similar arguments give a new and consistent explanation for quantum teleportation \cite{Bennett}.

The SI implies free spin-0 particles will never have \textit{Zitterbewegung}. This may have implications for the problems of self-energy and renormalization of fundamental particles \cite{Barut}. 

Finally, if the SI is valid for the KGE, than it should be valid in other areas of quantum mechanics and in quantum field theory. Future papers will explore this possibility.
%%%%%%%%%%%%%%%%%%%%%%%%%%%%%%%%%%%%%%%%%%%%%
\section{Acknowledgements} 
I thank Eleanor G. Rieffel, Kenneth B. Wharton, and Eugene D. Commins for many useful conversations.
%%%%%%%%%%%%%%%%%%%%%%%%%%%%%%%%%%%%%%%%%%%%%
 
%%%%%%%%%%%%%%%%%%%%%%%%%%%%%%%%%%%%%%%%%%%%%
\end{document}